\documentclass[twocolumn,aps,prl,showpacs]{revtex4}
\usepackage{amsmath,graphicx,color,amssymb,bm}

\begin{document}
\title{Band dispersion and electronic lifetimes in crystalline organic 
 semiconductors}

\author{S. Ciuchi$^{1}$ and S. Fratini$^{2}$} 
\affiliation{$^1$Istituto dei Sistemi Complessi CNR, CNISM and Dipartimento di Fisica,
Universit\`a dell'Aquila, via Vetoio, I-67100 Coppito-L'Aquila, Italy\\
$^2$ Institut N\'eel-CNRS and Universit\'e Joseph Fourier,Bo\^ite
Postale 166, F-38042 Grenoble Cedex 9, France}

\date{\today}

\begin{abstract}  
The consequences of several microscopic interactions 
on the photoemission spectra of crystalline organic semiconductors (OSC)
are studied theoretically. It is argued that
their relative 
roles can be disentangled by analyzing both their  temperature and
their momentum/energy dependence.  Our analysis shows that 
the polaronic thermal band narrowing, that is the foundation
of most theories of electrical transport in OSC, is
inconsistent in the range of microscopic parameters appropriate 
for these materials. An alternative scenario is proposed to explain
the experimental trends. 
\end{abstract} 

\maketitle

\paragraph{Introduction.}
It is now becoming possible to measure Angle-Resolved
photoemission (ARPES) spectra of organic
semiconductors (OSC) with increasingly high accuracy, 
revealing intrinsic properties of
conduction electrons in these materials. Clear energy-momentum dispersions,
indicative of the formation of well-defined electronic bands, have
been reported in pentacene monolayers \cite{Kakuta}, crystalline
films \cite{Hatch,Hatch09}   as well as  rubrene single
crystals \cite{Machida}. Since photoemission probes the very nature of
electronic excitations, the measured spectra (in particular, their
deviations from non-interacting Bloch bands)  
provide direct information on the interactions affecting the electron
motion. 
Ultimately, such information could help establishing a proper
microscopic model 
for the electron dynamics  in these materials.
For this procedure to be viable, however, one must be 
able to provide reliable predictions from the theoretical side,
identifying characteristic signatures of the 
different microscopic phenomena at work. 
Only then the comparison with the experimental ARPES spectra can be
effectively used to sort out the relevant microscopic interactions in
crystalline OSC.

In this work, we consider the effect of several microscopic
interactions that have been considered to play a role in crystalline OSC:
(i) the coupling of electrons to intra-molecular vibrations; 
(ii) dynamical disorder originating from thermal inter-molecular motions;
(iii) static disorder, of both chemical and structural origin, 
giving rise to spatial variations of 
the  molecular site energies and 
(iv) details of the electronic structure, most importantly 
the features arising from the non-equivalence of 
the molecules in the unit cell.
On general grounds, these interactions can affect either the 
band dispersion, the
electronic lifetimes (i.e. respectively the position and 
width of the ARPES peaks), or both. 
In order to disentangle all these effects experimentally, it is
important to characterize their  temperature dependences:
for example,   (i, ii) can in principle depend on temperature,
via the  thermal  changes in  the phonon population while  
(iii,iv) should be essentially temperature
independent, being determined by the structural characteristics of the sample.
A similar 
assessment can be made concerning their momentum/energy dependence, 
i.e.  how different interactions 
affect distinct regions of the electronic spectrum.  
Our aim here is to clarify and quantify these aspects
by performing a reliable
calculation of the electronic spectral function in a controlled
microscopic model.

\paragraph{Model.}

We consider the following  Hamiltonian, 
\begin{equation}
  \label{eq:Htot}
  H= H^{(0)} + H^{(i)}+H^{(ii)}+H^{(iii)}+H^{(iv)}+H^{(vib)}
\end{equation}
\begin{eqnarray}
H^{(0)}&=&-J \sum_j (c^+_j  c_{j+1} + c^+_{j+1} c_{j}) \label{eq:tb}\\
H^{(i)}&=& \sum_j g \; c^+_j  c_{j} X_j  \label{eq:Hol}\\
H^{(ii)}&=& -\sum_j f(u_j-u_{j+1}) \; (c^+_j  c_{j+1} + c^+_{j+1}
c_{j})  \label{eq:SSH} \\
H^{(iii)} &=& \sum_j  \epsilon_j c^+_j  c_{j} \;\;\; ; \;\;   
H^{(iv)}=  \delta \sum_j  (-1)^j c^+_j  c_{j} \label{eq:dimer}\\
H^{(vib)}&=& \sum_j     \frac{M \Omega_{0}^2X_j^2}{2}
+ \frac{P_j^2}{2M}  +  \frac{m \omega_{0}^2u_j^2}{2}
+ \frac{p_j^2}{2m}.     \label{eq:ph}
\end{eqnarray}
$H^{(0)}$ describes the tight binding motion of holes in a HOMO band
that we take to be one dimensional for simplicity, 
reflecting the marked
anisotropy that is commonly observed in crystalline OSC.
$H^{(i)}$ and $H^{(ii)}$ are respectively the ``Holstein''
interaction  with intra-molecular deformations $X_j$ of
frequency $\Omega_{0}$, with coupling strength $g$, 
and the ``Peierls'' or ``Su-Schrieffer-Heeger''
(SSH) interaction with molecular displacements $u_j$ of frequency
$\omega_{0}$. These two interactions modulate respectively the molecular 
energies and the inter-molecular transfer integrals.
For the latter we assume a linear dependence 
on the inter-molecular distance, $f(u_j-u_{j+1})=J\alpha_{SSH}
(u_j-u_{j+1}) $, being $\alpha_{SSH}$ the coupling strength.
$H^{(iii)}$ describes disordered  site energies
$\epsilon_j$ obeying a gaussian distribution of fixed width
$\Delta$, and 
$H^{(iv)}$ an alternating potential $\pm \delta$ reflecting a non-equivalence
of the sites in the unit cell, as is commonly observed in OSC.  
$H^{(vib)}$ is the Hamiltonian for free harmonic vibrations.
In the following we set $\hbar=k_B=1$
and express the strength of 
the electron-vibration interactions by introducing
the dimensionless parameters 
$\alpha_{H}=g/\Omega_{0}$
and  $\lambda_{SSH}=\alpha_{SSH}^2 J/(2 M\omega^2_{0})$. 
Typical bandwidts in organic semiconductors are 
$W\sim 0.5eV$, corresponding to $J\sim 0.125 eV$ in our one-dimensional model.
The microscopic  parameters  that can be found in the literature 
are  $\omega_0\simeq (0.05-0.1)J$ and $\Omega_0\simeq (1-2)J$ with 
$\lambda_{SSH}\simeq 0.15-0.3$ and $\alpha_{H}^2\simeq 0.2-0.5$,
and $\delta=(0.1-0.3)J$, $\Delta=(0.1-0.3)J$
\cite{Hannewald,Coropceanu02,DaSilva,Girlando10,Sanchez-Carrera,Hatch,Hatch09,
Machida,Kakuta,Yoshida,Kalb}.

\paragraph{Method.}

Owing to the low frequency of the inter-molecular
vibrations,  that stems from the large molecular mass, we have
both $\omega_0\ll J$ and $\omega_0 \lesssim T$ in the relevant
temperature range. In this regime 
the displacements $u_j$ in $H^{(ii)}$ can
be treated as a random field obeying a gaussian distribution,
causing a statistical disorder in the inter-molecular transfer
integrals \cite{note}.  
The solution of all the terms in Eq. (\ref{eq:Htot}) but $H^{(i)}$  therefore
reduces to a problem of non-interacting electrons in the presence of 
both ``site-diagonal'' and ``off-diagonal'' disorder, that can be
efficiently treated with the method described in Ref. \cite{Fratini09}.   
Instead, the full quantum nature of the
intra-molecular vibrations
must be retained  in $H^{(i)}$, because  $T\ll
\Omega_{0}$.  
To this aim we devise a method of solution that provides reliable
results in the  
case $\Omega_0\gtrsim J$, which includes  
the moderate vibrational frequency regime of interest here, 
$\Omega_0\simeq (1-2)J$.
As we show below, this treatment of the intra-molecular
electron-vibration interactions can be straightforwardly
implemented in the disordered environment provided by the remaining
terms in the Hamiltonian.

The Green's 
function of the problem can be defined in the site representation as
\begin{equation}
G_{i,j}=\int \Pi_l du_l \Pi_m d\epsilon_m  P({\bf u})
P_{dis}(\bm{\epsilon}) G_{i,j}({\bf u},\bm{\epsilon})
\label{eq:Gav}
\end{equation}
where 
$P({\bf u})\propto \Pi_l \exp({-u_l^2/2\sigma^2}) $ 
is the statistical distribution of local displacements
${\bf u}= \{u_l\}$, with $\sigma^2=[2M\omega_0 \tanh(\omega_0/2T)]^{-1}$
and $P_{dis}(\bm {\epsilon})\propto\Pi_m \exp (-\epsilon^2_m/2\Delta^2)$ is
the  distribution of 
local energies
$\bm{\epsilon}=\{\epsilon_m\}$. 
For a given set of inter-molecular deformations and disorder variables
the Green's function 
of the electronic problem $H_{el}=H-H^{(i)}$  in the absence of the
Holstein term  can be obtained 
as an inversion of a tri-diagonal matrix in the site representation:
$G_{i,j}({\bf u},\bm { \epsilon})=[(\omega - H_{el})^{-1}]_{i,j}$ with 
\cite{Fratini09}
\begin{eqnarray}
\label{eq:matrix}
\omega - H_{el}=\left( \begin{array}{cccc}
a_0^{(0)} &  b_0 & 0 &...\\
 b_0 & a^{(0)}_1 & b_1 & ...\\
...&...&...&...\\
...&0& b_{N-1} & a^{(0)}_N \end{array} \right).
\end{eqnarray}
The interaction term $H^{(i)}$ is subsequently included 
at a local level. In practice, we consider
a thermalized Holstein impurity at each site $i$, whose
Green's function is 
defined as
\begin{equation}
  \label{eq:thermG}
G^{(i)}_{loc}(\omega)= \frac{1}{Z_{H}}\left[\sum_{n=0}^\infty
\frac{e^{-n \Omega_{0}/T}}
{\mathcal{G}_{i,i}^{-1}(\omega) - \Sigma_{H,i}^{em,(n)} - \Sigma_{H,i}^{em,(n)}} \right]
\end{equation}
with $Z_H=\sum_{n=0}^\infty e^{-n \Omega_{0}/T}$  [we drop
the indices $({\bf u},\bm { \epsilon})$ in the following].  Here
$\mathcal{G}_{i,i}$ is the  local propagator obtained from the inversion
of the matrix Eq. (\ref{eq:matrix})
and $\Sigma_{H,i}^{em,(n)}$, $\Sigma_{H,i}^{em,(n)}$ are the 
local  emission/absorption self-energies in the n-phonon 
propagator  of the  impurity. These can be expressed as
continued fractions \cite{depolarone}:
\begin{widetext}
\begin{eqnarray}
\label{eq:CFE}
\Sigma_{H,i}^{em,(n)}(\omega)&=&(n+1)g^2/\displaystyle\mathcal{G}_{i,i}^{-1}
[\omega-\Omega_0]-\displaystyle (n+2)g^2
    /\displaystyle\mathcal{G}_{i,i}^{-1}[\omega-2\Omega_0]- \displaystyle\ldots\;\;\; ; \;\; n\ge 0  
\\\Sigma_{H,i}^{abs,(n)}(\omega)&=&n
g^2/\displaystyle\mathcal{G}_{i,i}^{-1}[\omega+\Omega_0]-(n-1)g^2/ 
\displaystyle\mathcal{G}_{i,i}^{-1}[\omega+2\Omega_0]-
\displaystyle
\displaystyle\ldots  \;\;\; ; \;\;n\ge 1 ,
\end{eqnarray}
\end{widetext} 
the latter fraction obviously ending at $n$-th stage.
To evaluate the  Green's function 
$G_{i,j}({\bf u},\bm{\epsilon})$
of  the full lattice problem, we now
invert Eq. (\ref{eq:matrix}) with the 
replacement $a^{(0)}_i=\omega - [H_{el}]_{i,i} \longrightarrow 
a_i=\omega - [H_{el}]_{i,i}-\Sigma_{H,i}$
where
\begin{equation}
\label{eq:self}
\Sigma_{H,i}=(\mathcal{G}_{i,i})^{-1}-(G^{(i)}_{loc})^{-1}
\end{equation}
is the local self-energy of the Holstein-impurity problem at each site $i$
[the off-diagonal terms $b_i=J+f(u_i-u_{i+1})$ are kept unchanged]. 
This treatment of the Holstein electron-vibration coupling 
has formally the same structure as 
Dynamical Mean Field Theory \cite{depolarone}, except for the absence
of a self-consistency loop. It reduces to the MA$^{(0)}$ approximation
of Refs. \cite{Berciu06,Berciu10} at $T=0$ and, as in 
MA$^{(0)}$, it is most
appropriate in the non-adiabatic regime, $\Omega_{0}\gtrsim J$.

The above Eqs. (\ref{eq:thermG})-(\ref{eq:self}) solve the 
Holstein interaction problem  for any given configuration of the 
diagonal and off-diagonal disorder. The 
solution of the
full Hamiltonian Eq. (\ref{eq:Htot}) 
is obtained from Eq. (\ref{eq:Gav}) upon averaging over $50000$
realizations of disorder variables  on a chain of $N=512$ sites. 
The spectral function is $A(k,\omega)=-\frac{1}{\pi}
Im G(k,\omega)$, where  $G(k,\omega)=\frac{1}{N^2}\sum_{i,j} e^{i k a(i-j)}
G_{i,j}(\omega)$ is the Green's function in momentum space.

\paragraph{Results.}
\begin{figure}
\includegraphics[width=4.7cm]{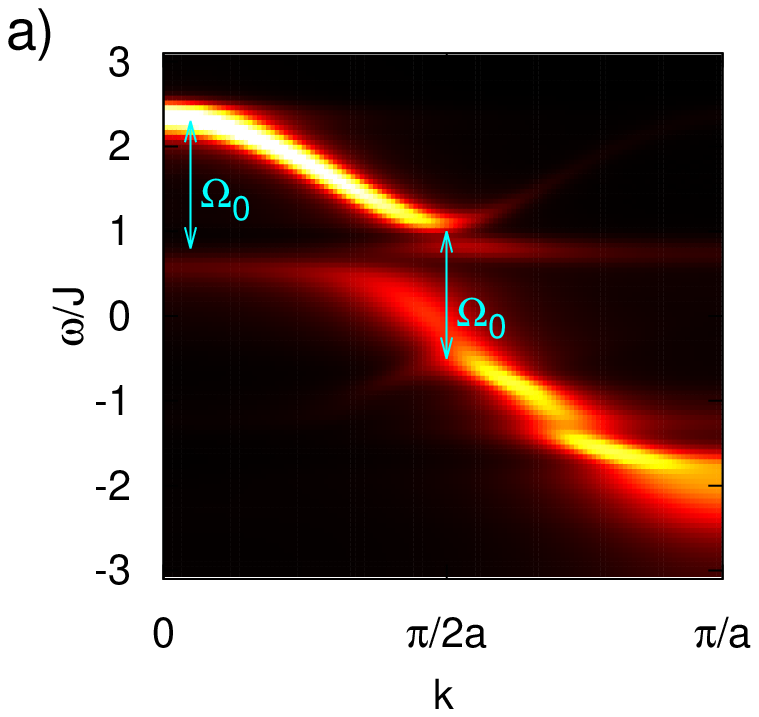} 
\includegraphics[width=3.8cm]{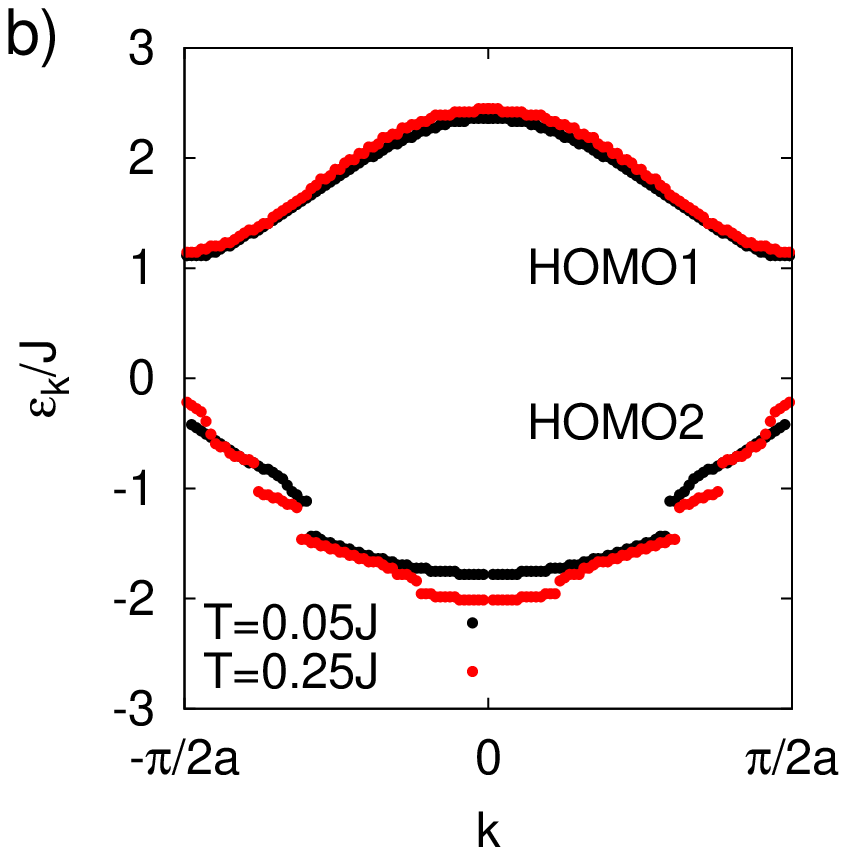}
  \centering
    \caption{(a) Spectral function $A(k,\omega)$ 
    for a hole in a  HOMO band, calculated for $J=0.125
eV$, $\lambda_{SSH}=0.2$, $\omega_0=0.05J$, $\alpha_{H}^2=0.33$,
$\Omega_0=1.5J$,  $\delta=0.3J$, $\Delta=0.2J$ at $T=0.05J$.
(b) Position of the maxima, tracking the renormalized band
dispersion $\varepsilon_k$ at low and room temperature (reduced Brillouin zone).
\label{fig:feticcio}}
\end{figure}

Fig. \ref{fig:feticcio}a shows a  photoemission 
spectrum calculated with the above procedure 
for holes in a one-dimensional  HOMO
band, ideally representing the direction of maximum conduction in an 
organic crystal, for a representative choice of parameters
[slight changes in the parameter values within the range indicated
after Eqs. (1)-(6)
do not appreciably 
modify the scenario].
To understand the results
it is instructive to analyze the different terms in the Hamiltonian
separately. 
We start from the interaction with the intra-molecular vibrations, 
$H^{(i)}$,  that is responsible for the most prominent
features observed in the  spectrum.
As is well known, in the case of isolated molecules this term 
gives rise to ``shakeoff'' satellites of the molecular levels, 
appearing  at multiples of the vibrational
frequency $\Omega_0$ \cite{Coropceanu02}.   
The number of visible satellites is set by the coupling strength, 
$N_{vib}=\alpha_H^2$, and is indicative of
the amount of vibrational quanta constituting the polaronic deformation. 

In a crystalline environment the situation is more complex.
The molecular picture is only recovered when the
electronic band dispersion 
is small compared to 
the vibration frequency, $W \ll \Omega_0$.
This however does not apply to organic crystals, where 
despite the arguably narrow bandwidths as compared to inorganic
semiconductors,  $W$ is still the largest energy
scale in the problem. Being $W>\Omega_0$, 
the periodic overtones characteristic of  molecular spectra  
are replaced 
by cuts in the band dispersion.
These features are 
analogous to the 
``kinks'' that are commonly observed in the photoemission spectra 
of solids with sizable electron-boson coupling. 
 They add a vibrational fine
structure to the spectrum \cite{Yamane}
without much affecting the overall bandwidth 
\cite{depolarone,Hohenadler,Berciu06}.
This phenomenon 
is clearly visible in Fig. \ref{fig:feticcio}a, where the cosine
dispersion starting from the top of the band is cut out by phonon
resonances at $\Omega_0$ and $2\Omega_0$
(indicated by arrows).  In the case $\Omega_0\sim (1-2)J$  that is
relevant to OSC, 
the interaction with intra-molecular
vibrations  effectively
splits the HOMO dispersion into two main subbands  separated by a 
sizable direct gap, as shown in Fig. \ref{fig:feticcio}b.
The gap opens  up where the dispersion crosses the first vibrational
cut, which  falls
accidentally around $k\sim \pi/2$.
This result can explain the large separation 
between the two HOMO subbands of pentacene 
\cite{Hatch09,Hatch}, 
that is one order of magnitude larger than that predicted by ab-initio
calculations on accounts of the structural non-equivalence in the unit cell 
\cite{Yoshida,notedelta}.

\begin{figure}
  \centering
\includegraphics[width=4cm]{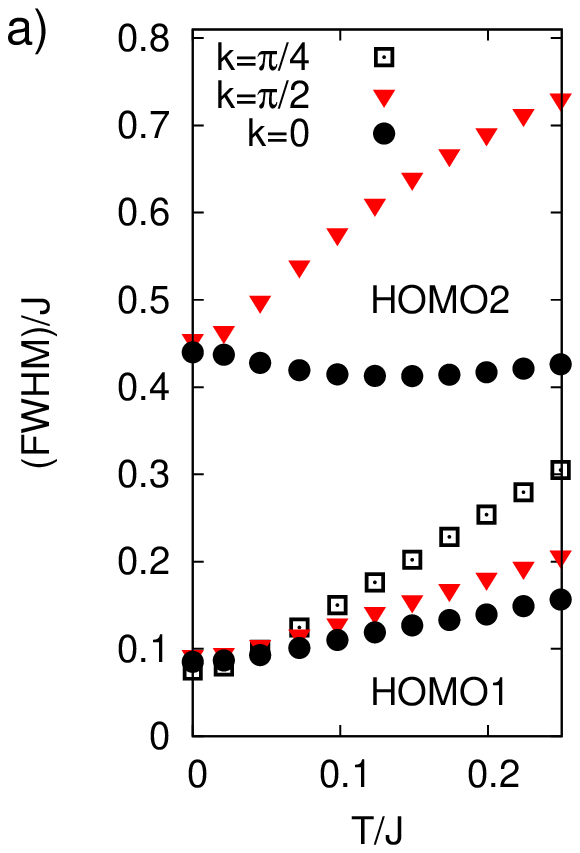}
\includegraphics[width=4cm]{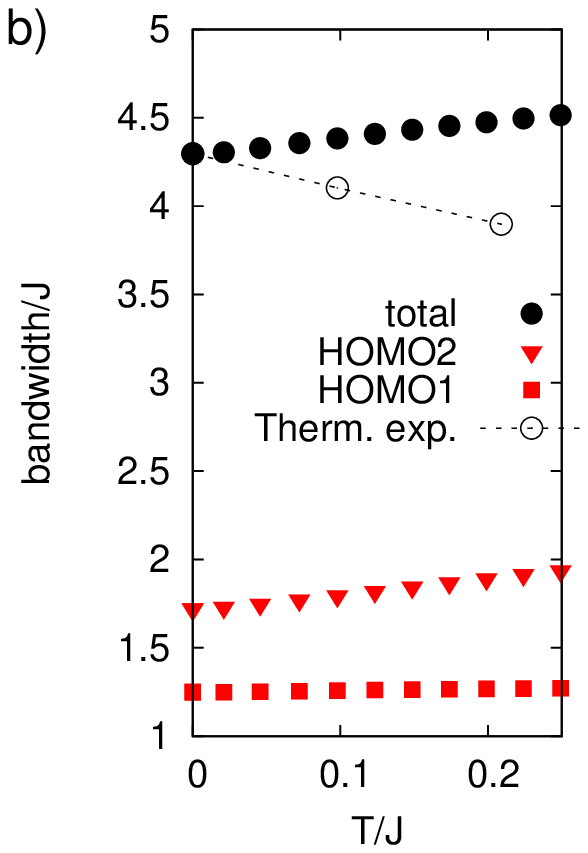}
    \caption{(a) Temperature dependence    of the spectral
        linewidths at different points in the BZ, for the same
        parameters as in Fig. 1.    
(b) Widths of the individual subbands and total bandwidth. Open
circles are obtained including the lattice thermal expansion
calculated in Ref. \cite{Masino}
(the bandwidth is normalized to the  zero-temperature value of $J$).
\label{fig:analysis}}
\end{figure}

Fig. \ref{fig:analysis}a reports 
the lifetimes for states at different points of the Brillouin zone
(BZ). The large
difference between the HOMO1 and HOMO2 branches, that was also observed
in Ref. \cite{Hatch},  is an additional distinctive feature of
the intra-molecular interaction $H^{(i)}$: 
according to the arguments given above, the HOMO1 band lies 
by construction
below the threshold for the emission of a
vibrational quantum,  $|\varepsilon_{k}-\varepsilon_{k=0}|< \Omega_0$,
and therefore the electronic lifetime there is mostly
insensitive to the effects of intra-molecular vibrations
\cite{depolarone,Berciu06}.   
Vibronic scattering
processes are instead allowed in the second subband, where they
cause a much larger line broadening, with linewidths of the order
of the electronic transfer rate itself.
Since only the HOMO1 states near the band edge can be thermally
populated or doped in a field-effect device, we argue that
scattering from high-frequency  vibrations 
should not play a predominant
role in the transport mechanism of OSC.

The effects related to  $H^{(i)}$ are essentially
temperature independent, because the considered vibrations cannot be
thermally excited in the relevant range $T\ll \Omega_0$. 
The temperature dependence observed in Fig. \ref{fig:analysis}a
therefore arises from 
other microscopic mechanisms present in the model.
While the static disorder, $H^{(iii)}$, can be excluded
(it adds a constant  broadening $\propto \Delta$ to the spectral
lines),
the scattering from slow
inter-molecular lattice vibrations in $H^{(ii)}$
is indeed strongly temperature-dependent: because the lattice motions are
thermally excited  for $T\gtrsim \omega_0/2$, 
the linewidths increase linearly with temperature, which is indeed observed in 
the HOMO1 sector of Fig. \ref{fig:analysis}a.
Different slopes arise  at different points of the BZ because
the scattering from inter-molecular lattice vibrations  
is minimum at the band edges 
\cite{Fratini09}.

Finally, Fig. 2b illustrates the temperature dependence of the calculated
electronic bandwidths.  
The total HOMO width (black dots)
exhibits a moderate  increase with
temperature, that is entirely caused by the coupling to low-frequency
inter-molecular vibrations: 
the thermal fluctuations of the lattice structure 
lead to an average  increase of the transfer
integrals given by $J_{av}^2 =J^2+\sigma_J^2$, with 
$\sigma_J^2=4 \lambda_{SSH} J T$ for $T\gtrsim \omega_0/2$
\cite{Gosar,Fratini09,TroisiDOS,Sanchez-Carrera}. 
The coupling to low-frequency 
{\it intra}-molecular vibrations, not considered here, 
would also lead to an analogous effect \cite{HoHu}.
Importantly,  this trend is inverted if one takes into account 
the large thermal
expansion coefficient characteristic of organic crystals (open circles in
Fig. 2b)  \cite{Haas}:  upon
heating,
the lattice expansion leads to an effective  reduction of the ``bare'' 
transfer integral $J$ entering  in Eq. (2)  \cite{Masino}.
The resulting bandwidth reduction  quantitatively
agrees with the thermal band narrowing reported in  pentacene
\cite{Koch,Hatch}. This effect should not be confused with the 
the phenomenon of {\it polaronic band narrowing} \cite{HolsteinII}, that 
is the foundation of popular 
theories of electrical transport in crystalline OSC, but 
 is inconsistent with the parameter range relevant to these materials.

In  standard polaron theories  \cite{HolsteinII,Silbey,Hannewald},  that 
build on the molecular limit  $W \ll \Omega_0$, 
the dressing of charge carriers by {\it fast} molecular vibrations causes
a renormalization of the transfer integrals and a consequent 
reduction of the bandwidths,
which is exponential in the phonon number
$N_{vib}$.
Upon increasing the temperature the ``polaronic'' band 
further shrinks due to the increased 
population of molecular vibrations, following $W\to W e^{-N_{vib}(T)}$, with
$N_{vib}(T)= \alpha_H^2 [1+2n_B(T)]$ and $n_B(T)$ the Bose factor.  
For a measurable effect to be observed at room temperature, however,
one would require $\Omega_0\lesssim 1000K\simeq 90 meV$. Such value is
considerably smaller than the bandwidth and therefore contradicts the
molecular limit assumption. 

In the regime of moderate coupling strengths and $\Omega_0<W$  of interest
for OSC, it is the very hypothesis of an
exponential renormalization that fails \cite{Berciu06,Hohenadler,HoHu}. 
A polaronic band can still be identified 
--- it is nothing but the HOMO1 band of Fig.1 --- 
but its width is now of the order of 
the vibration frequency $\Omega_0$, and not governed by the 
textbook exponential  relation. 
If we now focus on the entire HOMO dispersion, 
the absence of polaronic band narrowing is even more evident: 
with the present parameters 
one would predict a $30\%$
polaronic reduction of the bandwidth from the non-interacting value
$W=4J$, which is clearly not observed in Fig. 1 (we show in the suplementary 
EPAPS file that this is not a drawback of the present method).
The physical reason for the failure of the conventional picture
is that the molecular
degrees of freedom  are not fast enough to rearrange during the electron
motion, and therefore cannot provide the instantaneous renormalization implied
in the molecular limit.


\paragraph{Concluding remarks.}

The calculations presented here show that the photoemission spectra
of crystalline organic semiconductors can differ significantly from the
non-interacting band picture,
reflecting a sizable coupling of the conduction electrons with other degrees
of freedom in the system.
The most salient features in the spectra 
come from the interaction with high-frequency
intra-molecular vibrations, that causes 
the opening of apparent gaps in the electronic dispersion
 at multiples of the vibrational 
energy $\Omega_0$.  
However,  such high-frequency modes  
are unable to efficiently scatter the quasiparticle 
states at the lowest binding energies.
For this reason, the origin of the temperature dependence of 
the mobility in these materials cannot be explained by polaronic
effects but should rather be seeked in the interaction with
inter-molecular vibrational modes of much lower frequency, whose 
temperature-dependent scattering rates are directly accessible in 
ARPES experiments.

\acknowledgments

We thank A. Girlando for enlightening discussions.


\begin{thebibliography}{30}
\bibitem{notedelta} 
The large HOMO1/HOMO2 gap in Fig. \ref{fig:feticcio} 
remains unchanged if we set  $\delta=0$. The effect of the
structural term $H^{(iv)}$ reduces to the small splitting  
visible  at $k=\pi/2$ around $\omega\simeq J$  
and minor features in the HOMO2 band.
\end{thebibliography}
\end{document}